\documentclass{iaus}
\usepackage{graphicx}

\def\aa{\textit{A\&A}\ }
\def\annrev{\textit{ARA\&A}\ }
\def\apj{\textit{ApJ}\ }
\def\nat{\textit{Nature}\ }
\def\lsim{\mathrel{\rlap{\lower 4pt \hbox{\hskip 1pt $\sim$}}\raise 1pt
\hbox {$<$}}} 
\def\gsim{\mathrel{\rlap{\lower 4pt \hbox{\hskip 1pt $\sim$}}\raise 1pt
\hbox {$>$}}}
\newcommand{\Msun}{M_\odot}
\newcommand{\Mms}{M_{\rm MS}}
\newcommand{\Nifs}{$^{56}$Ni}
\newcommand{\Mni}{M{\rm (^{56}Ni)}}
\newcommand{\Ed}{\dot{E}_{\rm dep}}
\newcommand{\Edep}{\dot{E}_{\rm dep,51}}

\title[Gamma-Ray Bursts and Extremely Metal-Poor Stars] 
{The Connection between Gamma-Ray Bursts and Extremely Metal-Poor
Stars as Nucleosynthetic Probes of the Early Universe}

\author[Ken'ichi Nomoto {\textit et al.}]   
{K.~Nomoto$^{1,2}$, N.~Tominaga$^1$, M.~Tanaka$^1$, K.~Maeda$^2$, 
 \and H.~Umeda$^1$}

\affiliation{$^1$Department of Astronomy, University of Tokyo, Bunkyo-ku, Tokyo
113-0033, Japan \\ email: {\tt nomoto@astron.s.u-tokyo.ac.jp} \\[\affilskip]
$^2$Institute for the Physics and Mathematics of the Universe,
University of Tokyo, Kashiwa, Chiba 277-8582, Japan}

\pubyear{2008}
\volume{250}  
\pagerange{119--126}
\jname{Massive Stars as Cosmic Engines}
\editors{F.\ Bresolin, P.A.\ Crowther, \& J.\ Puls, eds.}
\begin{document}

\maketitle

\begin{abstract}

The connection between the long GRBs and Type Ic Supernovae (SNe) has
revealed the interesting diversity: (i) GRB-SNe, (ii) Non-GRB
Hypernovae (HNe), (iii) X-Ray Flash (XRF)-SNe, and (iv) Non-SN GRBs
(or dark HNe).  We show that nucleosynthetic properties found in the
above diversity are connected to the variation of the abundance
patterns of extremely-metal-poor (EMP) stars, such as the excess of C,
Co, Zn relative to Fe.  We explain such a connection in a unified
manner as nucleosynthesis of hyper-aspherical (jet-induced) explosions
of Pop III core-collapse SNe.  We show that (1) the explosions with
large energy deposition rate, $\Ed$, are observed as GRB-HNe and their
yields can explain the abundances of normal EMP stars, and (2) the
explosions with small $\Ed$ are observed as GRBs without bright SNe
and can be responsible for the formation of the C-rich EMP (CEMP) and
the hyper metal-poor (HMP) stars.  We thus propose that GRB-HNe and
the Non-SN GRBs (dark HNe) belong to a continuous series of BH-forming
massive stellar deaths with the relativistic jets of different $\Ed$.

\end{abstract}

\firstsection 
\section{Introduction}

\begin{figure}[t]
\begin{center}
\includegraphics[width=10.0cm]{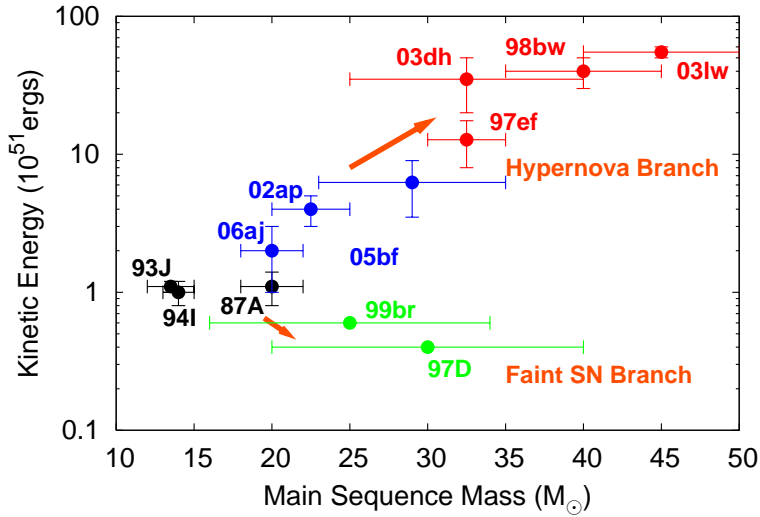}
\end{center}
\caption{
The kinetic explosion energy $E$ as a
function of the main sequence mass $M$ of the progenitors for several
supernovae/hypernovae. Hypernovae are the SNe with $E_{51}>10$. 
}
\label{figME}
\end{figure}

Among the most interesting recent developments in the study of
supernovae (SNe) is the establishment of the Gamma-Ray Burst
(GRB)-Supernova Connection (\cite{woo06}).  Three GRB-associated SNe
have been observed so far: GRB~980425/ SN~1998bw 
(\cite{galama1998,iwa98}), GRB~030329/SN~2003dh (\cite{sta03,hjo03}),
and GRB~031203/SN~2003lw (\cite{mal04}).  They are all very energetic
supernovae, whose kinetic energy $E$ exceeds $10^{52}$\,erg, more than
10 times the kinetic energy of normal core-collapse SNe.  In the
present paper, we use the term 'Hypernova (HN)' to describe such a
hyper-energetic supernova with $E_{51} = E/10^{51}$ erg $\gsim 10$
(Fig.\ref{figME}; \cite{nom04,nomoto2006}).  The above three SNe
associated with GRBs are called ``GRB-HNe''.

In contrast, ``non-SN GRBs'' (or {\sl dark HNe}) have also been
discovered (GRBs 060605 and 060614) (\cite{fyn06,gal06,del06,geh06}).
Upper limits to brightness of the possible SNe are about 100 times
fainter than SN~1998bw.  These correspond to upper limits to the
ejected \Nifs\ mass of $\Mni\sim 10^{-3}\Msun$ (see, e.g., 
\cite{nom06b} for prediction).

Such hypernovae and GRBs are also likely to be hyper-aspherical
explosions induced by relativistic jet(s) as suggested from
photometric and spectroscopic observations.

We calculate nucleosynthesis in such hyper-energetic and
hyper-aspherical explosions and find that resultant abundance features
show some important differences from normal supernova explosions (e.g., 
\cite{mae02,mae03,tom07,tom08}).  
We show that such features can explain the peculiar abundance patterns
observed in the extremely metal-poor (EMP), and hyper-metal-poor (HMP)
halo stars (e.g., \cite{hill2005, beers2005}).  This approach would
lead to identifying the First Stars in the Universe, which is one of
the important challenges of the current astronomy.

\begin{figure}[t]
\begin{center}
\includegraphics[width=8.5cm]{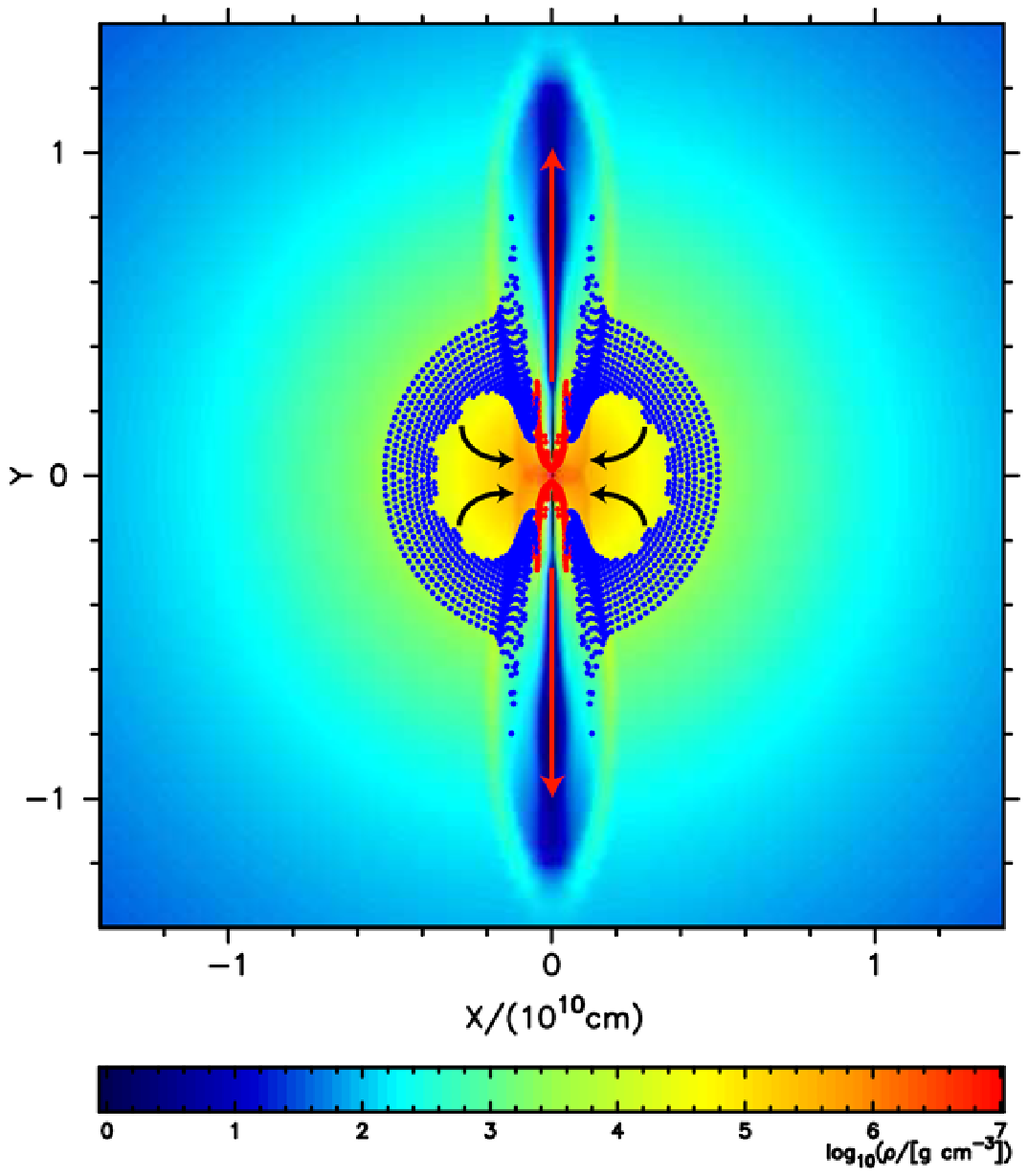}
\caption{The density structure of the 40 $\Msun$ Pop III star
 explosion model of $\Edep=15$ at 1 sec after the start of the jet
 injection. The jets penetrate the stellar mantle ({\it red arrows}) and
 material falls on the plane perpendicular to the jets ({\it black arrows}). 
 The dots represent ejected Lagrangian elements dominated by Fe 
 (\Nifs, {\it red}) and by O ({\it blue}).}
\label{fig:fallback}
\end{center}
\end{figure}

\begin{figure}[t]
\begin{center}
\includegraphics[width=8.5cm]{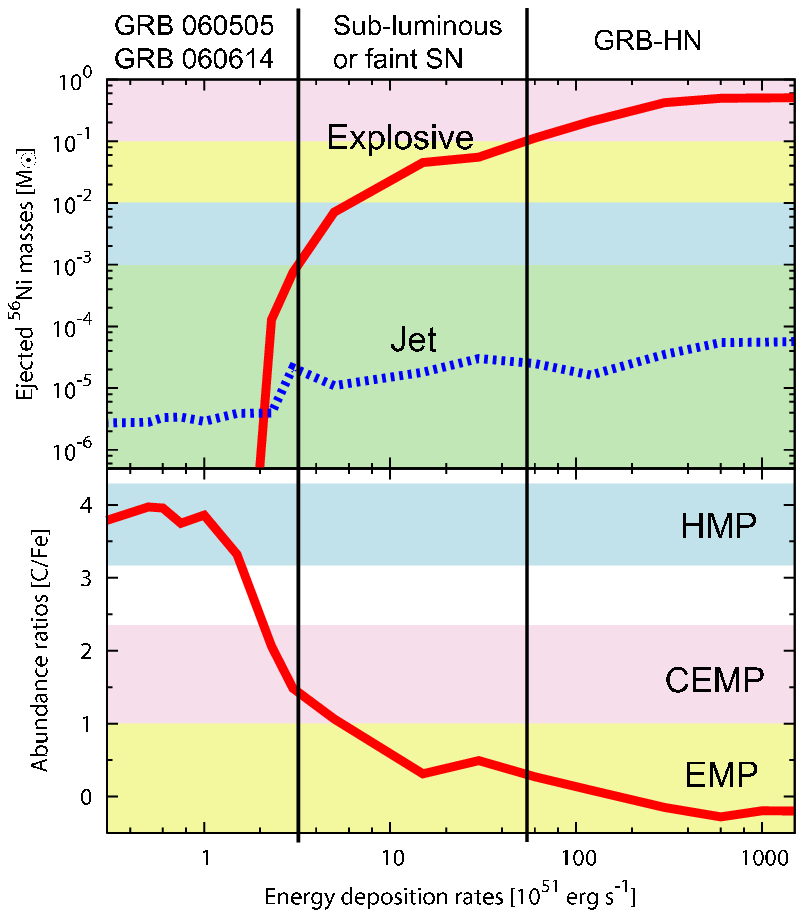}
\caption{{\it Upper}: the ejected \Nifs\ mass ({\it red}: 
 explosive nucleosynthesis products, {\it blue}: the jet contribution)
 as a function of the energy deposition rate. The background color
 shows the corresponding SNe ({\it red}: GRB-HNe, {\it yellow}:
 sub-luminous SNe, {\it blue}: faint SNe, {\it green}: GRBs~060505 and
 060614).  Vertical lines divide the resulting SNe according to their
 brightness.  {\it Lower}: the dependence of abundance ratio [C/Fe]
 on the energy deposition rate. The background color shows the
 corresponding metal-poor stars ({\it yellow}: EMP, {\it red}: CEMP,
 {\it blue}: HMP stars).}
\label{fig:EdotNi}
\end{center}
\end{figure}

\section{Nucleosynthesis in Jet-Induced Explosions}

We calculate hydrodynamics and nucleosynthesis of the explosions
induced by relativistic jets ({\sl jet-induced explosions})
(Fig.~\ref{fig:fallback}) (\cite{tom07,tom08}).  For the $40\Msun$ stars
(\cite{ume05,tominaga2006}), the jets are injected at a radius $R_0 \sim
900$ km (corresponding to an enclosed mass of $M_0 \sim 1.4 \Msun$).
The most important parameter in our models is the energy deposition
rate $\Ed$.  We investigate the dependence of nucleosynthesis outcome
on $\Ed$ for a range of $\Edep\equiv\Ed/10^{51}{\rm
ergs\,s^{-1}}=0.3-1500$.  The diversity of $\Ed$ is consistent with
the wide range of the observed isotropic equivalent $\gamma$-ray
energies and timescales of GRBs (\cite{ama06} and references therein).
Variations of activities of the central engines, possibly
corresponding to different rotational velocities or magnetic fields,
may well produce the variation of $\Ed$.

The upper panel of Figure~\ref{fig:EdotNi} shows the dependence of the
ejected $\Mni$ on $\Ed$.  Generally, higher $\Ed$ leads to the
synthesis of larger $\Mni$ in explosive nucleosynthesis because of
higher post-shock densities and temperatures (e.g.,
\cite{mae03,nag06}).  If $\Edep\gsim60$, we obtain $\Mni$
$\gsim0.1\Msun$, which is consistent with the brightness of GRB-HNe.
Some C+O core materials are ejected along the jet-direction, but a
large amount of materials along the equatorial plane fall back
(Fig.~\ref{fig:fallback}).

For $\Edep\gsim60$, the remnant mass is initially $M_{\rm rem}^{\rm
start}\sim1.5\Msun$ and grows as materials are accreted from the
equatorial plane (Fig.~\ref{fig:fallback}).  The final BH mass is
generally larger for smaller $\Ed$.  The final BH masses range from
$M_{\rm BH}=10.8\Msun$ for $\Edep=60$ to $M_{\rm BH}=5.5\Msun$ for
$\Edep=1500$, which are consistent with the observed masses of
stellar-mass BHs (\cite{bai98}).  The model with $\Edep=300$
synthesizes $\Mni\sim0.4\Msun$ and the final mass of BH left after the
explosion is $M_{\rm BH}=6.4\Msun$.

For low energy deposition rates ($\Edep<3$), in contrast, the ejected
\Nifs\ masses ($\Mni<10^{-3}\Msun$) are smaller than the upper
limits for GRBs~060505 and 060614. 

\begin{figure}[t]
\begin{center}
\includegraphics[width=10cm]{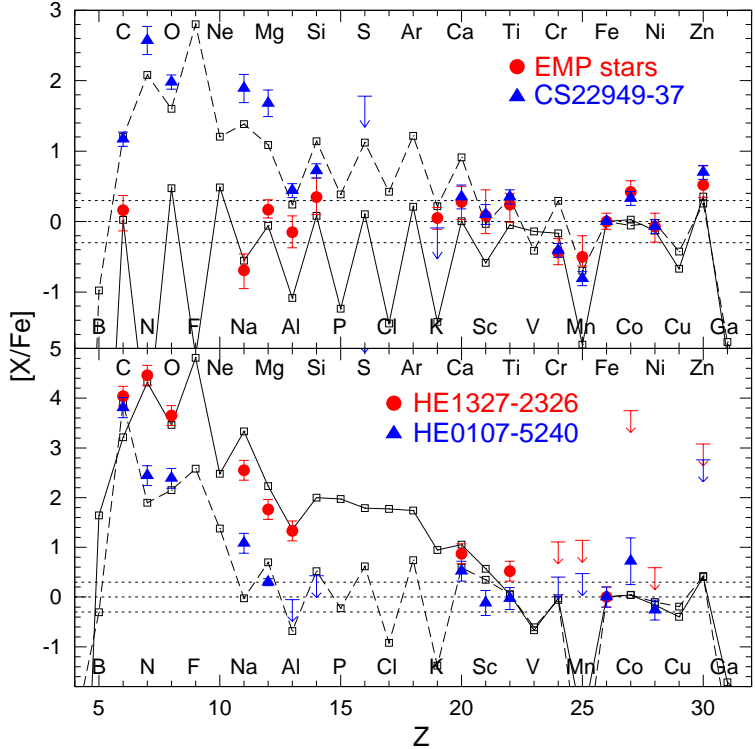}
\caption{
 A comparison of the abundance patterns between the metal-poor 
 stars and our models. 
 {\it Upper}: typical EMP ({\it red dots}, \cite{cayrel2004}) and 
 CEMP ({\it blue triangles}, CS~22949--37, \cite{dep02}) stars and models
 with $\Edep=120$ ({\it solid line}) and $=3.0$ ({\it dashed line}).
 {\it Lower}: HMP stars:
 HE~1327--2326, ({\it red dots}, e.g., \cite{fre05}), 
 and HE~0107--5240, ({\it blue triangles}, \cite{chr02,bes05}) and models 
 with $\Edep=1.5$ ({\it solid line}) and $=0.5$ ({\it dashed line}).
\label{fig:HMP}}
\end{center}
\end{figure}

If the explosion is viewed from the jet direction, we would observe
GRB without SN re-brightening. This may be the situation for
GRBs~060505 and 060614.  In particular, for $\Edep<1.5$, \Nifs\ cannot
be synthesized explosively and the jet component of the Fe-peak
elements dominates the total yields (Fig.~\ref{fig:HMP}). The
models eject very little $\Mni$ ($\sim10^{-6}\Msun$).

For intermediate energy deposition rates ($3\lsim\Edep<60$), the
explosions eject $10^{-3}\Msun \lsim \Mni <0.1\Msun$ and the final BH
masses are $10.8\Msun\lsim M_{\rm BH}< 15.1\Msun$. The resulting SN is
faint ($\Mni <0.01\Msun$) or sub-luminous ($0.01\Msun \lsim \Mni
<0.1\Msun$).

\section{Abundance Patterns of Extremely Metal-Poor Stars}

Table 1 summarizes the abundance features of various EMP stars.  In
addition to HMP stars, we focus on the recently discovered first Ultra
Metal-Poor (UMP) star (\cite{nor07}) and the very peculiar Si-poor
star (\cite{coh07}).  Many of these EMP stars have high [Co/Fe],
suggesting the HN-connection.

\begin{figure}[t]
\begin{center}
\includegraphics[width=10cm]{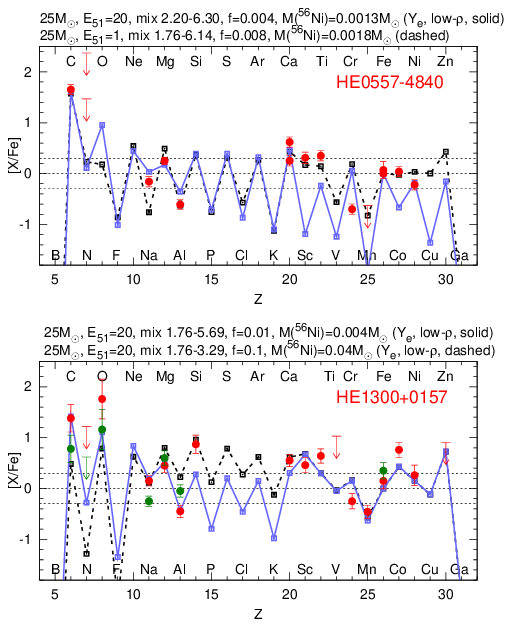}
\caption{Comparisons of the abundance patterns between the
 mixing-fallback models and the UMP star HE0557--4840 (upper:
 \cite{nor07}), and the CEP star HE1300+0157 (lower: \cite{fre07}).
}
\label{figHE0557}
\end{center}
\end{figure}

\subsection{C-rich Metal-Poor Stars (CEMP)}

The lower panel of Figure~\ref{fig:EdotNi} shows the dependence of
the abundance ratio [C/Fe] on $\Ed$.  Lower $\Ed$ yields larger
$M_{\rm BH}$ and thus larger [C/Fe], because the infall decreases the
amount of inner core material (Fe) relative to that of outer material
(C) (see also \cite{mae03}).  As in the case of $\Mni$, [C/Fe] changes
dramatically at $\Edep\sim3$.

The abundance patterns of the EMP stars are good indicators of SN
nucleosynthesis because the Galaxy was effectively unmixed at [Fe/H]
$< -3$ (e.g., \cite{tum06}). They are classified into three groups
according to [C/Fe]:

(1) [C/Fe] $\sim 0$, normal EMP stars ($-4<$ [Fe/H] $<-3$, e.g., 
    \cite{cayrel2004});

(2) [C/Fe] $\gsim+1$, Carbon-enhanced EMP (CEMP) stars ($-4<$ [Fe/H]
    $<-3$, e.g., CS~22949--37, \cite{dep02});

(3) [C/Fe] $\sim +4$, hyper metal-poor (HMP) stars ([Fe/H] $<-5$,
    e.g., HE~0107--5240, \cite{chr02,bes05}; HE~1327--2326,
    \cite{fre05}).

Figure \ref{fig:HMP} shows that the abundance patterns of the averaged
normal EMP stars, the CEMP star CS~22949--37, and the two HMP stars
(HE~0107--5240 and HE~1327--2326) are well-reproduced by models with
$\Edep=120$, 3.0, 1.5, and 0.5, respectively. The model for the normal
EMP stars ejects $\Mni\sim0.2\Msun$, i.e., a factor of 2 less than
SN~1998bw. On the other hand, the models for the CEMP and the HMP
stars eject $\Mni\sim8\times10^{-4}\Msun$ and $4\times 10^{-6}\Msun$,
respectively, which are always smaller than the upper limits for
GRBs~060505 and 060614.  The N/C ratio in the models for CS~22949--37
and HE~1327--2326 is enhanced by partial mixing between the He and H
layers during presupernova evolution (\cite{iwamoto2005}).

\subsection {UMP Star HE~0557--4840 and CEMP-no Star HE~1300+0157}

The abundance pattern of the first Ultra Metal-Poor (UMP) star
(HE~0557--4840: \cite{nor07}) is shown in Figure~\ref{figHE0557} and
compared with the HN ($E_{51}=20$) and SN ($E_{51}=1$) models of
the $25\Msun$ stars.  The Co/Fe ratio ([Co/Fe]$\sim0$) requires a high
energy explosion and the high [Sc/Ti] and [Ti/Fe] ratios require a
high-entropy explosion.  As shown in Figure~\ref{figHE0557} (upper), a
HN model with a ``low-density'' modification (\cite{tom07}) is in a
good agreement with the abundance pattern of HE~0557--4840. The model
indicates $\Mni\sim10^{-3}\Msun$ being similar to faint SN models for
CEMP stars.  The [Cr/Fe] ratio in the model is much higher than that
of HE~0557--4840.

The abundance pattern of the CEMP-no star (i.e., CEMP with no neutron
capture elements) HE~1300+0157 (\cite{fre07}) is shown in
Figure~\ref{figHE0557} (lower) and marginally reproduced by the
hypernova model with $\Mms=25\Msun$ and $E_{51}=20$.  The large [Co/Fe]
particularly requires the high explosion energy.

\begin{table}
  \caption{Metal-poor stars.}
  \label{tab:stars}
  \begin{tabular}{ccccccccc}
   \hline
    Name & [Fe/H] & Features & Reference \\
   \hline
   HE~0107--5240 & $-5.3$ & C-rich, Co-rich?, [Mg/Fe]$\sim0$ &
    \cite{chr02} \\
   HE~1327--2326 & $-5.5$ & C, O, Mg-rich & \cite{fre05,aok06} \\
   HE~0557--4840 & $-4.8$ & C, Ca, Sc, Ti-rich, [Co/Fe]$\sim0$
 & \cite{nor07} \\
   HE~1300+0157 & $-3.9$ & C, Si, Ca,Sc,Ti, Co-rich
 & \cite{fre07} \\
   HE~1424--0241 & $-4.0$ & Co,Mn-rich, Si,Ca,Cu-poor
 & \cite{coh07} \\
   CS~22949--37 & $-4.0$ & C,N,O,Mg,Co,Zn-rich
 & \cite{dep02} \\
   CS~29498--43 & $-3.5$ & C,N,O,Mg-rich, [Co/Fe]$\sim0$
 & \cite{aok04} \\
   BS~16934--002 & $-2.8$ & O,Mg-rich, C-poor
 & \cite{aok07} \\
   \hline
  \end{tabular}
\end{table}

\subsection {Si-Poor Star: HE~1424--0241}

\begin{figure}[t]
\begin{center}
\includegraphics[width=10cm]{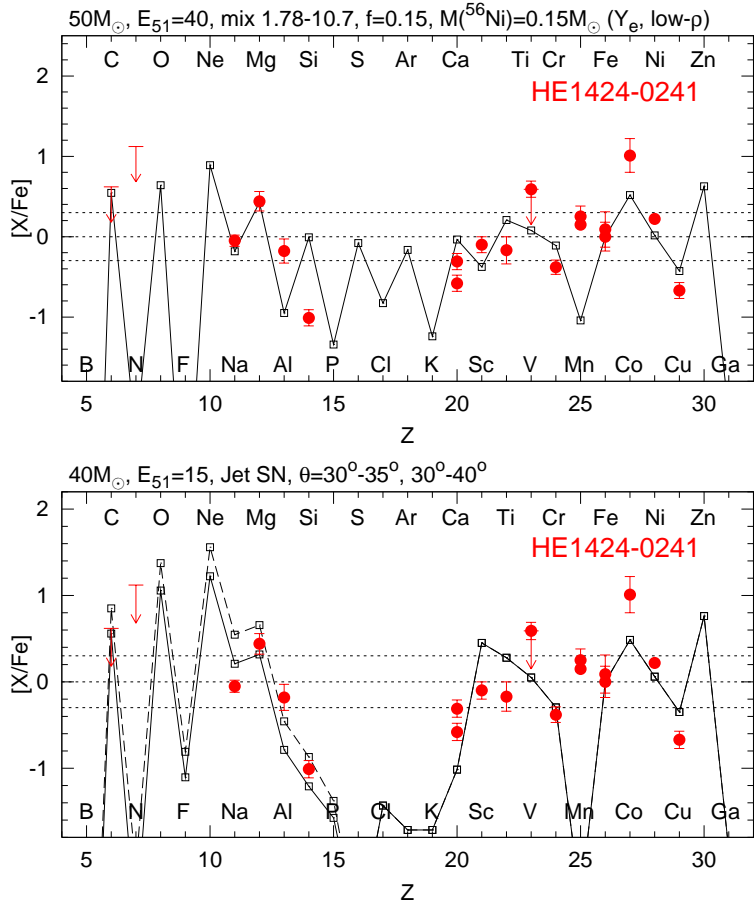}
\caption{Comparisons between the abundance patterns of HE1424--0241
 (\cite{coh07}) and the mixing-fallback model (upper), 
and the angle-delimited yields integrated over
 $30^\circ-40^\circ$ ({\it dashed line}) and $30^\circ-35^\circ$ ({\it
 solid line}) of the jet-induced SN model with $\dot{E}_{\rm
 dep}=1.2\times10^{53}~{\rm ergs~s^{-1}}$ (lower).
}
\label{figHE1424}
\end{center}
\end{figure}

The very peculiar Si-poor abundance pattern of HE~1424--0241
(\cite{coh07}) is shown in Figure~\ref{figHE1424} (upper) and compared
with the model of $\Mms=50\Msun$ and $E_{51}=40$.  The high [Mg/Si]
ratio cannot be reproduced by this model.  The peculiar abundance
pattern of HE~1424--0241 is a challenge to the explosion models.

The angle-delimited yield provides a possibility to explain the high
[Mg/Si] and normal [Mg/Fe].  Figure~\ref{figHE1424} (lower) shows that
the yields integrated over $30^\circ-40^\circ$ or $30^\circ-35^\circ$
reproduce the abundance pattern of HE~1424--0241. The yields consist
of Mg in the inner region and Fe in the outer region.

Thus the most difficult pattern can be reproduced by the angular
dependence of the yield.  The high [Mg/Si] and normal [Mg/Fe] are
realized if the heavy elements penetrate into the stellar mantle and
expand laterally (i.e., the duration of the jet injection is long) and
if Mg along the equatorial plane is not accreted onto the central
region (i.e., $\Ed$ is large).

\section{Concluding Remarks}

We show that (1) the explosions with large energy deposition rate,
$\Ed$, are observed as GRB-HNe and their yields can explain the
abundances of normal EMP stars, and (2) the explosions with small
$\Ed$ are observed as GRBs without bright SNe and can be responsible
for the formation of the CEMP and the HMP stars.  We thus propose that
GRB-HNe and GRBs without bright SNe belong to a continuous series of
BH-forming massive stellar deaths with the relativistic jets of
different $\Ed$.
The very peculiar Si-poor EMP star can also be explained by the
angle-delimited yield.

\end{document}